# Cesium Enhances Long-Term Stability of Lead Bromide Perovskite-Based Solar Cells

Michael Kulbak[a,†], Satyajit Gupta[a,†], Nir Kedem[a], Igal Levine[a], Tatyana Bendikov[b], Gary Hodes[a,*] and David Cahen[a,*]

[a]Department of Materials & Interfaces, Weizmann Institute of Science, Rehovot, 76100, Israel.

[b]Department of Chemical Research Support, Weizmann Institute of Science, Rehovot, 76100, Israel.

**ABSTRACT:** Direct comparison between perovskite-structured hybrid organic-inorganic - methyl ammonium lead bromide ($MAPbBr_3$) and all-inorganic cesium lead bromide ($CsPbBr_3$), allows identifying possible fundamental differences in their structural, thermal and electronic characteristics. Both materials possess a similar direct optical band-gap, but $CsPbBr_3$ demonstrates a higher thermal stability than $MAPbBr_3$. In order to compare device properties we fabricated solar cells, with similarly synthesized $MAPbBr_3$ or $CsPbBr_3$, over mesoporous titania scaffolds. Both cell types demonstrated comparable photovoltaic performances under AM1.5 illumination, reaching power conversion efficiencies of ~6 % with a poly-aryl amine-based derivative as hole transport material. Further analysis shows that Cs-based devices are as efficient as, and more stable than methyl ammonium-based ones, after aging (storing the cells for 2 weeks in a dry (relative humidity 15-20%) air atmosphere in the dark) for 2 weeks, under constant illumination (at maximum power), and under electron beam irradiation.

**KEYWORDS:** Perovskite solar cells, $CsPbBr_3$, stability.

Solar cells based on hybrid organic inorganic perovskites (HOIPs) with the generic structural formula $AMX_3$ (where A is an organic cation, M is the metal center and X is a halide) have shown rapidly increasing efficiencies[1,2] if methyl ammonium (MA) or formamidinium (FA)[3] is the organic monovalent cation in the 'A' site (which has a permanent dipole moment), M=lead ($Pb^{2+}$) and X = a monovalent halide anion.[4,5,6] Most efforts focus on iodide-based materials, which show the highest efficiencies. Cells made with the higher band gap bromide-based perovskites generate an open circuit voltage ($V_{OC}$) of up to ~1.5 V.[7,8,9] Such cells are of interest for possible use in tandem or spectral splitting systems and for photoelectrochemistry to generate energy-storing chemicals by e.g., water splitting and carbon dioxide ($CO_2$) reduction, provided they are stable.

Recently, we showed that the organic cation can be replaced by cesium, $Cs^+$, to form cesium lead bromide ($CsPbBr_3$), with a completely inorganic perovskite (CIP) structure at standard temperature and pressure.[10] We found photovoltaic (PV) devices made with this material to yield efficiencies as high as those of analogous HOIP ones.[10] That result calls for a direct comparison of the all-inorganic to the methyl ammonium lead bromide ($MAPbBr_3$)-based cells both in terms of PV performance and stability, using perovskites that are prepared in the same manner. Here we report such comparison between the $Cs^+$- and MA-based lead-halide perovskites in terms of thermal properties, and the corresponding photovoltaic device performance and stability.

In our earlier work the active $CsPbBr_3$ perovskite layer of the cell was deposited in two steps and processing was carried out in ambient atmosphere,[10] while the $MAPbBr_3$ was deposited with a one-step process.[8] To be able to compare the materials and devices made with them, the two materials were prepared under as identical processing conditions as possible (using the two-step process) and the same holds for devices made with them, including materials and thickness of the electron transport material (ETM), mesoporous layer and hole transport material (HTM). The results of such comparison show that the all-inorganic material and device performance are more stable than those made with the HOIPs, under both operational and storage conditions.

The structural, electronic and thermal properties of $MAPbBr_3$ and $CsPbBr_3$ were characterized using X-ray diffraction (XRD), UV-visible spectroscopy, ultraviolet photoelectron spectroscopy (UPS), contact potential difference (CPD) measurements and thermogravimetric analysis (TGA). The XRD patterns of the $CsPbBr_3$ (Figure S1A) and $MAPbBr_3$ (Figure S1B), deposited on mesoporous titania (mp-$TiO_2$)-coated FTO slides correspond to previously-published patterns of $CsPbBr_3$ and $MAPbBr_3$.[11,12] Optical absorption/bandgap measurements also agree with previously-published data with a direct bandgap for $MAPbBr_3$ of 2.32 eV[8] and a direct bandgap of 2.36 eV possibly preceded by an indirect gap of ~2.3 eV for $CsPbBr_3$ (the $CsPbBr_3$ spectrum and bandgap were discussed in Reference 10).

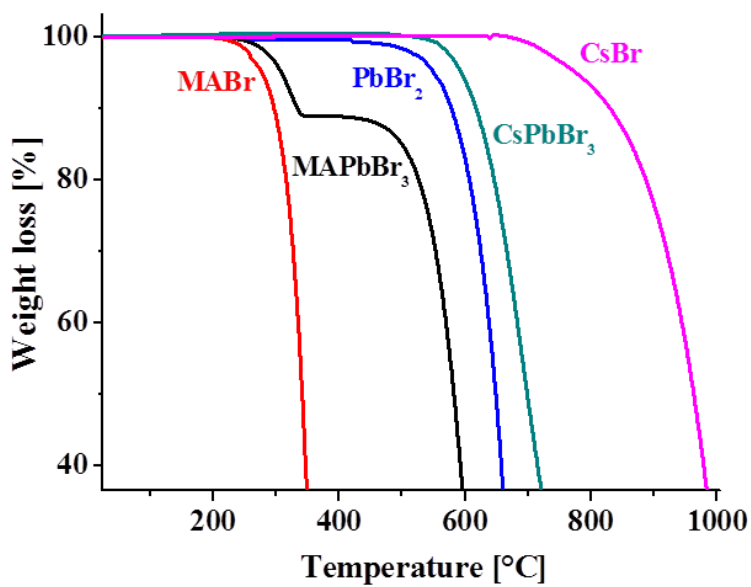

**Figure 1.** Thermogravimetric analyses of methyl ammonium bromide (MABr), methyl ammonium lead bromide (MAPbBr$_3$), lead bromide (PbBr$_2$), cesium lead bromide (CsPbBr$_3$) and cesium bromide (CsBr), showing the higher thermal stability of the inorganic perovskite compared to the hybrid organic-inorganic perovskite.

Values of work function and ionization energy/valence band energies of the two perovskites deposited on FTO\dense (d)-TiO$_2$ were measured by UPS. For CsPbBr$_3$ the values were 3.95 eV and 5.75 eV, respectively and for MAPbBr$_3$, 4.15 eV and 6.1 eV. The work functions of the materials were also obtained from CPD (using the Kelvin-Probe technique) in vacuum (Table S1) and agree well with the UPS measurements. These studies show that the differences in surface energetics between the materials are small.

A significant difference between the materials was observed in terms of real-time thermal stability, as shown by TGA analysis. Figure 1 shows a higher degradation onset temperature of CsPbBr$_3$ (~580°C) than of MAPbBr$_3$ (first onset at ~220°C). MAPbBr$_3$ displays degradation in stages suggesting defragmentation of the 'labile' organic cation (pristine MABr shows an onset of degradation at ~200°C, as shown in Figure 1; that of CsBr is ~650 ºC). This observation agrees with results from earlier studies on $MAPbX_3$-based systems.[13,14] It is possible that at the first stage (~250°C; 14% loss of the overall weight), $MAPbBr_3$ partially degrades into its constituents, as methylamine and some hydrogen bromide (HBr), while at the second stage (~432°C, 75% of the total loss) there is a complete degradation of the perovskite, indicated by a sharp weight loss. While, $CsPbBr_3$ shows a sharp, single-stage degradation at ~580°C, without any lower temperature features, indicating that the material is stable close to that temperature, due to the very different temperature stabilities of MABr and CsBr (and, in general, MAX vs. CsX; X = halide). In fact, $CsPbBr_3$ is somewhat more thermally stable than $PbBr_2$ itself.

To study the photovoltaic behavior of $MAPbBr_3$ and $CsPbBr_3$, devices with a configuration FTO/d-$TiO_2$/mp-$TiO_2$/perovskite/HTM/Au were fabricated, using PTAA (HOMO ~5.2 eV) as HTM. The Cs- and MA-based devices demonstrated comparable performance under AM 1.5 illumination and the J-V results for the best performing cells (in the dark and illuminated) are given in Figure 2. The distributions in various photovoltaic parameters for $MAPbBr_3$- and $CsPbBr_3$-based cells are shown in Figures S2 and S3, respectively. Average values of the various parameters calculated from these cells are as follows: For $MAPbBr_3$-based cells: $V_{OC}$ = 1.37 V, $J_{SC}$ = 5.9 mA/cm$^2$, fill factor = 71%, efficiency =5.8%. For $CsPbBr_3$-based cells: $V_{OC}$ = 1.26 V, $J_{SC}$ = 6.2 mA/cm$^2$, fill factor = 74 % and efficiency = 5.8%. The $CsPbBr_3$ cells gave a somewhat lower $V_{OC}$ but this was compensated by a higher $J_{SC}$ and fill factor.

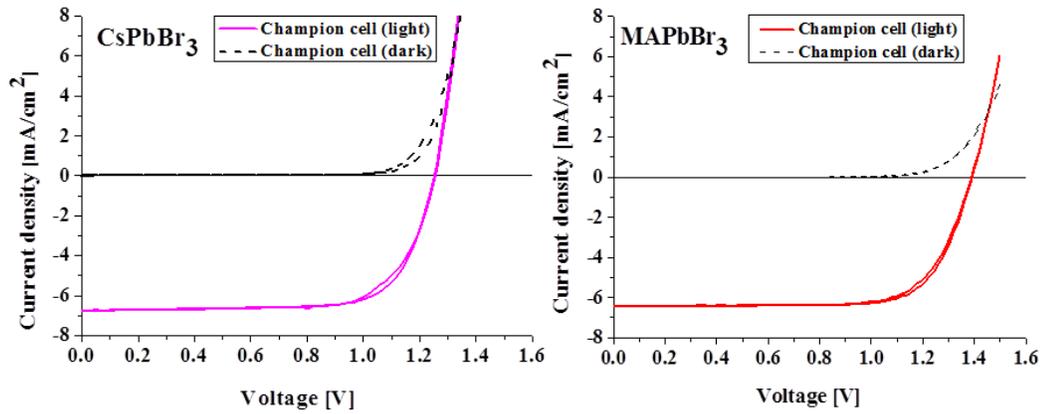

| Scan direction | $V_{OC}$ (Volt) | $J_{SC}$ (mA/cm$^2$) | FF (%) | PCE (%) |
|---|---|---|---|---|
| FWD | 1.25 | 6.7 | 73 | 6.2 |
| REV | 1.25 | 6.7 | 72 | 6.1 |

| Scan direction | $V_{OC}$ (Volt) | $J_{SC}$ (mA/cm$^2$) | FF (%) | PCE (%) |
|---|---|---|---|---|
| FWD | 1.38 | 6.4 | 73 | 6.5 |
| REV | 1.4 | 6.4 | 74 | 6.6 |

**Figure 2.** J-V characteristics of the best performing CsPbBr$_3$- and MAPbBr$_3$-based cells in the dark and under illumination, demonstrating comparable device performances and tabulated values of their PV parameters (bottom). [PCE: power conversion efficiency; FWD: forward; REV: reverse]

The lower $V_{OC}$ of the CsPbBr$_3$-based cells is of particular interest. In principle, it could be due to the deeper valence band of the MAPbBr$_3$, 6.1 eV compared to 5.75 eV for the CsPbBr$_3$, although both are much deeper than the (separately measured) valence band edge of the HTM (5.2 eV) and thus, based on a simple, and probably unrealistic, consideration of energy band alignments, no difference is expected for this reason This $V_{OC}$ difference is presently under investigation.

The operational stabilities of the cells were monitored by measuring the photocurrent densities at an applied bias close to the initial maximum power point ($V_{mp}$ ~1.04 V for MAPbBr$_3$ and ~1 V for CsPbBr$_3$) as a function of time for both cell types (Figure 3). During a 5 h illumination period, the MAPbBr$_3$ cell shows a strong decay (~55% loss, compared to the maximum value) in photocurrent density as a function of time, while CsPbBr$_3$ shows a much slower and smaller decay (~13%) in the photocurrent density in the same time frame. Note that all cells in this study were *not encapsulated*.

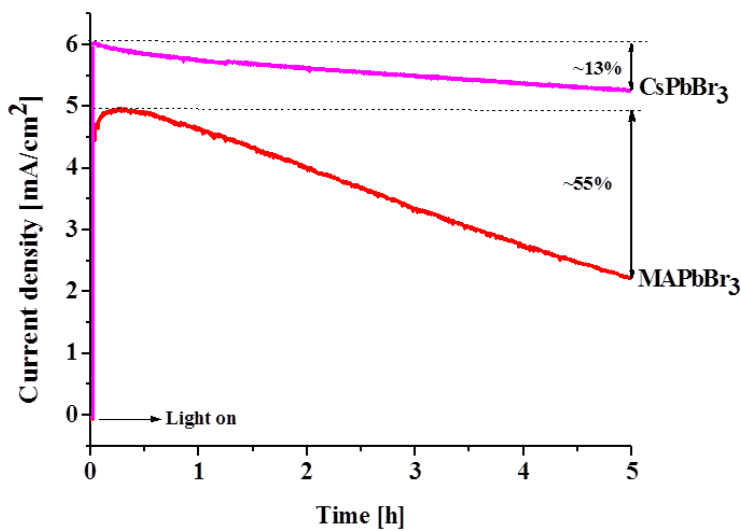

**Figure 3.** Current density measured at an applied bias close to the initial maximum power point vs time under 100 mW/cm$^2$ AM1.5 illumination for MAPbBr$_3$– and CsPbBr$_3$–based cells.

Another significant difference in device parameters between MAPbBr$_3$ and CsPbBr$_3$ was observed during aging studies, presented in Figure 4. The measurements were carried out in ambient air under relative humidity (RH) of 60-70 %, every couple of days for two weeks. Between measurements, the devices were kept in a dry air atmosphere (in the dark) with a RH of ~15-20 %. MAPbBr$_3$-based devices showed a steady decay in all device parameters, leading to an average loss of ~85% in efficiency, ~25 % in open circuit voltage, ~71 % in current density and ~35 % in the fill factor, while CsPbBr$_3$-based cells showed no significant decay (J-V curves of the devices are shown in Figure S4). One possible reason is the much higher volatility of MABr compared to CsBr; decomposition of the perovskites with water vapor results in MABr that can gradually volatilize away, while this happens much slower, if at all, with CsBr. Thus, while liquid water can both decompose and remove (some of) the decomposition products of both perovskites rather rapidly, water vapor is expected to have a much smaller effect on the Cs than on the MA perovskite. It may also be that the polar organic MA cation makes MAPbBr$_3$ more hydrophilic in character than CsPbBr$_3$ and thus allows water molecules to permeate faster through the edges of the devices, increasing the decomposition rate. In any case, higher device stability is critical for long-term practical device applications.

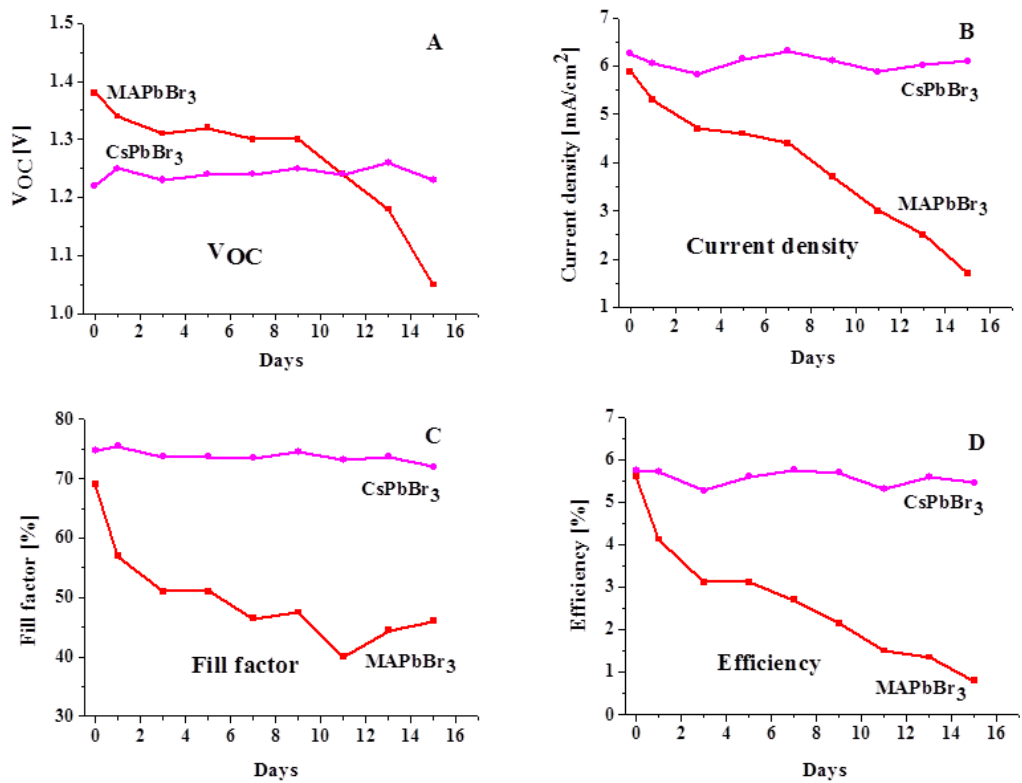

**Figure 4.** Aging analysis of MAPbBr$_3$ and CsPbBr$_3$ cells. Figures show the cell parameters **(A)** V$_{OC}$, **(B)** J$_{SC}$ **(C)** fill factor and **(D)** efficiency, as a function of time, demonstrating the much greater stability of CsPbBr$_3$-based cells with aging.

The devices were further analyzed using electron beam-induced current (EBIC) analysis. Figure 5 (A) and (B) show cross-sectional scanning electron microscopy images of the devices. The top row (A) is for a CsPbBr$_3$-based device, while the bottom row (B) is for a MAPbBr$_3$ one. The extreme left images of the collage are secondary electron (SE) ones (marked as 'SE image'). Sequential scanning images using EBIC analysis are displayed from left to right (marked as 'scan 1' to 'scan 5' in the images of the collage). In brief, EBIC uses an electron beam to act as a light source equivalent, generating electron-hole pairs in the junction area as depicted in Figure S5. These pairs separate into free carriers, which are collected at the contacts. The current originating from the charge collection is observed in real time and a current collection efficiency image can be drawn. In a mesoporous structure, a fixed collection efficiency is expected throughout the device due to the short collection distance. The EBIC signal in $CsPbBr_3$-based cells (Figure 5A) is stable from the FTO/d-$TiO_2$ interface through the mesoporous and capping perovskite layer. There is no apparent loss in charge collection when scanning the same cross sectional area multiple times, suggesting $CsPbBr_3$ remains stable and does not degrade under the electron beam. For the $MAPbBr_3$-based cells, the first EBIC image (Figure 5B) indicates efficient charge collection at the FTO/d-$TiO_2$ interface, similar to what we reported earlier.[15] Although the image drifts while repeating the EBIC scans, it is clear that the EBIC signal decays as the number of scans increases, meaning a decrease in collection efficiency. We attribute this to severe beam damage of the $MAPbBr_3$ due to extensive local heating at the beam point of entrance.

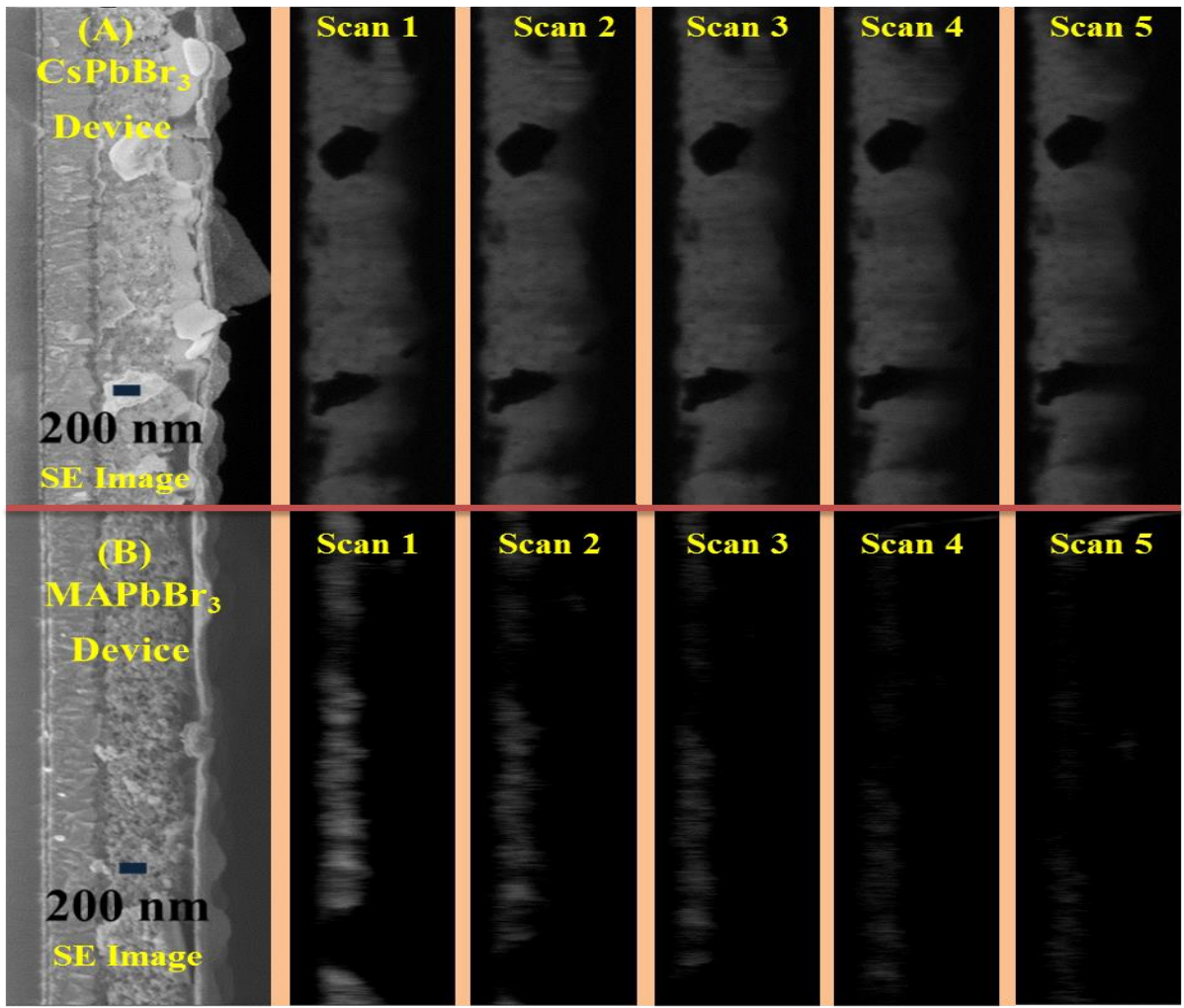

**Figure 5.** Repetitive sequential EBIC responses of cross-sections of **(A)** a CsPbBr$_3$ cell indicating a stable electron beam-induced current generation, whereas **(B)** the MAPbBr$_3$ cell, shows a steady decay in current under same conditions. ('Scan 1' to 'Scan 5' indicate the sequence of the scans).

It is likely that the thermal stability observed in TGA for CsPbBr$_3$ is associated with the total device stability, as well as its efficient collection efficiency under the electron beam, during EBIC analysis. We suggest that due to its relatively high temperature, single phase degradation process, the CsPbBr$_3$-based devices are less prone to environmental degradation and their life-time is prolonged compared to that of MAPbBr$_3$-based devices. As the inclusion of chloride in MAPbBr$_3$ was reported to increase its stability as well[7,16] we are now studying the effects of chloride addition to CsPbBr$_3$.

In conclusion, we have shown that replacing the common organic 'A' site of AMX$_3$ halide perovskites by an inorganic one forms a more thermally stable perovskite structure, as indicated by TGA analysis. The opto-electronic properties for both materials were investigated using UPS and CPD analysis together with J-V characterization, with specific emphasis on the stability of the devices. Devices fabricated with CsPbBr$_3$ demonstrated photovoltaic performance, comparable to that of MAPbBr$_3$–based ones but with much improved solar cell stability as shown in device aging studies. These results indicate that CsPbBr$_3$, a completely inorganic perovskite, is to be preferred as an ABX$_3$ absorber over MAPbBr$_3$ for long-term stable device operation. Further investigation is needed to understand the lower $V_{OC}$ (and to increase this parameter) and also the (modestly but reproducibly) higher $J_{SC}$ and fill factor obtained for cells with all-inorganic absorbers compared to cells with hybrid absorbers.

**EXPERIMENTAL SECTION**

**Device fabrication**

F-doped tin oxide (FTO) transparent conducting substrates (Xinyan Technology TCO-XY15) were cut and cleaned by sequential 15 minutes sonication in warm aqueous alconox solution, deionized water, acetone and ethanol, followed by drying under N$_2$ stream. After an oxygen plasma treatment for 10 minutes, a compact ~60 nm thin TiO$_2$ layer was applied to the clean substrate by spray pyrolysis of a 30 mM titanium diisopropoxide bis(acetylacetonate) (Sigma

Aldrich) solution in isopropanol using air as the carrier gas on a hot plate set to 450 °C, followed by a two-step annealing procedure at 160 °C and 500 °C, each for 1 hour in air.

A 450-nm-thick mesoporous $TiO_2$ scaffold was deposited by spin-coating a $TiO_2$ paste onto the dense $TiO_2$-coated substrates. A $TiO_2$ paste (DYESOL, DSL 18NR-T) and ethanol were mixed in a ratio of 2:7 by weight and sonicated until all the paste dissolved. The paste was spin-coated for 5 seconds at 500 rpm and 30 seconds at 2000 rpm, twice, followed by a two-step annealing procedure at 160 °C and 500 °C, each for 1 h in air.

The $MAPbBr_3$ and $CsPbBr_3$ films were prepared by a 2-step sequential deposition technique. For both cases, 1 M of $PbBr_2$ (Sigma Aldrich) in DMF was stirred on a hot plate at 75 °C for 20 minutes. It was then filtered using a 0.2 μm pore size PTFE filter and immediately used. The solution was kept at 75 °C during the spin-coating process. For preparation of the $CsPbBr_3$ film, the solution was spin-coated on pre-heated (75 °C) substrates for 30 seconds at 2500 rpm and was then dried on a hot plate at 70 °C for 30 minutes. After drying, the substrates were dipped for 10 minutes in a heated (50 °C) solution of 17 mg/ml CsBr (Sigma Aldrich) in methanol for 10 minutes, washed with 2-propanol, dried under $N_2$ stream and annealed for 10 minutes at 250 °C. For $MAPbBr_3$, the solution was spin-coated over un-heated substrates for 20 seconds at 3000 rpm, and was dried on a hot plate at 70 °C for 30 minutes. After drying, the substrates were dipped for 40 seconds in a 10 mg/ml MABr in isopropanol, washed with isopropanol and then dried under $N_2$ stream and annealed for 15 minutes at 100 °C. All procedures were carried out in an ambient atmosphere. poly[bis(4-phenyl)(2,4,6-trimethylphenyl)amine] (PTAA – Lumtec) was applied by spin-coating 5 seconds at 500 rpm followed by 40 seconds at 2000 rpm. For $CsPbBr_3$, the PTAA solution contained 15 mg in 1 mL of chlorobenzene, mixed with 7.5 μL of tert-butylpyridine (TBP) and 7.5 μL of 170 mg/mL

LiTFSI [bis(trifluoromethane)sulfonamide (in acetonitrile)], while for MAPbBr$_3$ the PTAA solution contained 30 mg in 1 mL of chlorobenzene, mixed with 15 μL of tert-butyl pyridine and 15 μL of 170 mg/mL LiTFSI. The samples were left overnight in the dark in dry air before ~100 nm gold contacts were thermally evaporated on the back through a shadow mask with 0.24 cm$^2$ rectangular holes.

**Characterization**

The thermogravimetric analyses was carried out using TA instruments, at a heating rate of 20°C/min (using alumina crucibles) under inert atmosphere. XRD measurements were conducted on a Rigaku ULTIMA III operated with a Cu anode at 40 kV and 40 mA. The measurements were taken using a Bragg-Brentano configuration through a 10 mm slit, a convergence Soller 5° slit and a 'Ni' filter. A Jasco V-570 spectrophotometer with an integrating sphere was used for measuring reflectance-corrected transmission. The J-V characteristics were measured with a Keithley 2400-LV SourceMeter and controlled with a Labview-based, in-house written program. A solar simulator (ScienceTech SF-150; with a 300 W Xenon short arc lamp from USHIO Inc., Japan) equipped with an AM1.5 filter and calibrated with a Si solar cell IXOLARTM High Efficiency SolarBIT (IXYS XOB17-12x1) was used for illumination. The devices were characterized through a 0.16 cm$^2$ mask. The J-V characteristics were taken after light soaking for 10 s at open circuit and at a scan rate of 0.06 V/s (unless otherwise stated). Ultraviolet Photoemission Spectroscopy (UPS) measurements were carried out using a Kratos AXIS ULTRA system, with a concentric hemispherical analyzer for photo excited electron detection. UPS was measured with a helium discharge lamp, using He I (21.22 eV) and He II (40.8 eV) radiation lines. The total energy resolution was better than 100 meV, as determined from the Fermi edge of gold (Au) reference. A Kelvin probe located in a controlled atmosphere station (McAllister Technical Services) was used to measure CPD between the probe and the sample surface, under a vacuum of ~$10^{-3}$ mbar. EBIC analysis was done in a Zeiss-Supra SEM using beam current of 5 pA and beam energy of 3 keV. Current was collected and amplified using Stanford Research Systems SR570 pre-amplifier. The device cross section was exposed by mechanical cleaving immediately (up to 2 min) before transferring the sample into the SEM vacuum chamber ($10^{-5}$ mbar).

## ASSOCIATED CONTENT

**Supporting Information**

Supporting information contains X-Ray diffractograms, work function values, statistics of cell parameters, J‐V curves taken during device aging and a schematic explaining the EBIC analysis. The supporting information is available free of charge on the ACS website at DOI: 10.1021/.

## AUTHOR INFORMATION

**Corresponding Authors**
*E-mail: gary.hodes@weizmann.ac.il.
*E-mail: david.cahen@weizmann.ac.il.

## AUTHOR CONTRIBUTIONS

†Contributed equally.

## NOTES

The authors declare no competing financial interest.

ACKNOWLEDGEMENTS

This research work was supported by the Israel Ministry of Science's Tashtiot, Israel-China and India-Israel programs, the Israel National Nano-initiative and the Sidney E. Frank Foundation through the Israel Science Foundation. D.C. holds the Sylvia and Rowland Schaefer Chair in Energy Research.

**REFERENCES**

(1) Jeon, N. J.; Noh, J. H.; Yang, W. S.; Kim, Y. C.; Ryu, S.; Seo, J.; Seok, S. I. Compositional Engineering of Perovskite Materials for High-Performance Solar Cells. *Nature* **2015**, *517* (7535), 476–480.

(2) Burschka, J.; Pellet, N.; Moon, S.-J.; Humphry-Baker, R.; Gao, P.; Nazeeruddin, M. K.; Grätzel, M. Sequential Deposition as a Route to High-Performance Perovskite-Sensitized Solar Cells. *Nature* **2013**, *499* (7458), 316–319.

(3) Eperon, G. E.; Stranks, S. D.; Menelaou, C.; Johnston, M. B.; Herz, L. M.; Snaith, H. J. Formamidinium Lead Trihalide: A Broadly Tunable Perovskite for Efficient Planar Heterojunction Solar Cells. *Energy Environ. Sci.* **2014**, *7* (3), 982.

(4) Berry, J.; Buonassisi, T.; Egger, D. A.; Hodes, G.; Kronik, L.; Loo, Y.-L.; Lubomirsky, I.; Marder, S. R.; Mastai, Y.; Miller, J. S.; et al. Hybrid Organic-Inorganic Perovskites (HOIPs): Opportunities and Challenges. *Adv. Mater.* **2015**, *27* (35), 5102–5112.

(5) Green, M. A.; Ho-Baillie, A.; Snaith, H. J. The Emergence of Perovskite Solar Cells. *Nat. Photonics* **2014**, *8* (7), 506–514.

(6) Hodes, G.; Cahen, D. Photovoltaics: Perovskite Cells Roll Forward. *Nat. Photonics* **2014**, *8* (2), 87–88.

(7) Edri, E.; Kirmayer, S.; Kulbak, M.; Hodes, G.; Cahen, D. Chloride Inclusion and Hole Transport Material Doping to Improve Methyl Ammonium Lead Bromide Perovskite-Based High Open-Circuit Voltage Solar Cells. *J. Phys. Chem. Lett.* **2014**, *5* (3), 429–433.

(8) Edri, E.; Kirmayer, S.; Cahen, D.; Hodes, G. High Open-Circuit Voltage Solar Cells Based on Organic–Inorganic Lead Bromide Perovskite. *J. Phys. Chem. Lett.* **2013**, *4* (6), 897–902.

(9) Ryu, S.; Noh, J. H.; Jeon, N. J.; Chan Kim, Y.; Yang, W. S.; Seo, J.; Seok, S. I. Voltage Output of Efficient Perovskite Solar Cells with High Open-Circuit Voltage and Fill Factor. *Energy Environ. Sci.* **2014**, *7* (8), 2614.

(10) Kulbak, M.; Cahen, D.; Hodes, G. How Important Is the Organic Part of Lead Halide Perovskite Photovoltaic Cells? Efficient $CsPbBr_3$ Cells. *J. Phys. Chem. Lett.* **2015**, *6* (13), 2452–2456.

(11) Tidhar, Y.; Edri, E.; Weissman, H.; Zohar, D.; Hodes, G.; Cahen, D.; Rybtchinski, B.; Kirmayer, S. Crystallization of Methyl Ammonium Lead Halide Perovskites: Implications for Photovoltaic Applications. *J. Am. Chem. Soc.* **2014**, *136* (38), 13249–13256.

(12) Stoumpos, C. C.; Malliakas, C. D.; Peters, J. A.; Liu, Z.; Sebastian, M.; Im, J.; Chasapis, T. C.; Wibowo, A. C.; Chung, D. Y.; Freeman, A. J.; et al. Crystal Growth of the Perovskite Semiconductor $CsPbBr_3$: A New Material for High-Energy Radiation Detection. *Cryst. Growth Des.* **2013**, *13* (7), 2722–2727.

(13) Liu, Y.; Yang, Z.; Cui, D.; Ren, X.; Sun, J.; Liu, X.; Zhang, J.; Wei, Q.; Fan, H.; Yu, F.; et al. Two-Inch-Sized Perovskite $CH_3NH_3PbX_3$ (X = Cl, Br, I) Crystals: Growth and Characterization. *Adv. Mater.* **2015**, *27* (35), 5176–5183.

(14) Stoumpos, C. C.; Malliakas, C. D.; Kanatzidis, M. G. Semiconducting Tin and Lead Iodide Perovskites with Organic Cations: Phase Transitions, High Mobilities, and Near-Infrared Photoluminescent Properties. *Inorg. Chem.* **2013**, *52* (15), 9019–9038.

(15) Kedem, N.; Brenner, T. M.; Kulbak, M.; Schaefer, N.; Levcenko, S.; Levine, I.; Abou-Ras, D.; Hodes, G.; Cahen, D. Light-Induced Increase of Electron Diffusion Length in a p–n Junction Type CH$_3$NH$_3$PbBr$_3$ Perovskite Solar Cell. *J. Phys. Chem. Lett.* **2015**, *6* (13), 2469–2476.

(16) Das, J.; Bhaskar Kanth Siram, R.; Cahen, D.; Rybtchinski, B.; Hodes, G. Thiophene-Modified Perylenediimide as Hole Transporting Material in Hybrid Lead Bromide Perovskite Solar Cells. *J Mater Chem A* **2015**, *3* (40), 20305–20312.

# Supporting Information

**Cesium Enhances Long-Term Stability of Lead Bromide Perovskite-Based Solar Cells**

Michael Kulbak[a,†], Satyajit Gupta[a,†], Nir Kedem[a], Igal Levine[a], Tatyana Bendikov[b], Gary Hodes[a,*] and David Cahen[a,*]

[a]Department of Materials & Interfaces, Weizmann Institute of Science, Rehovot, 76100, Israel.

[b]Department of Chemical Research Support, Weizmann Institute of Science, Rehovot, 76100, Israel.

**Corresponding Authors**
*E-mail: gary.hodes@weizmann.ac.il.
*E-mail: david.cahen@weizmann.ac.il.

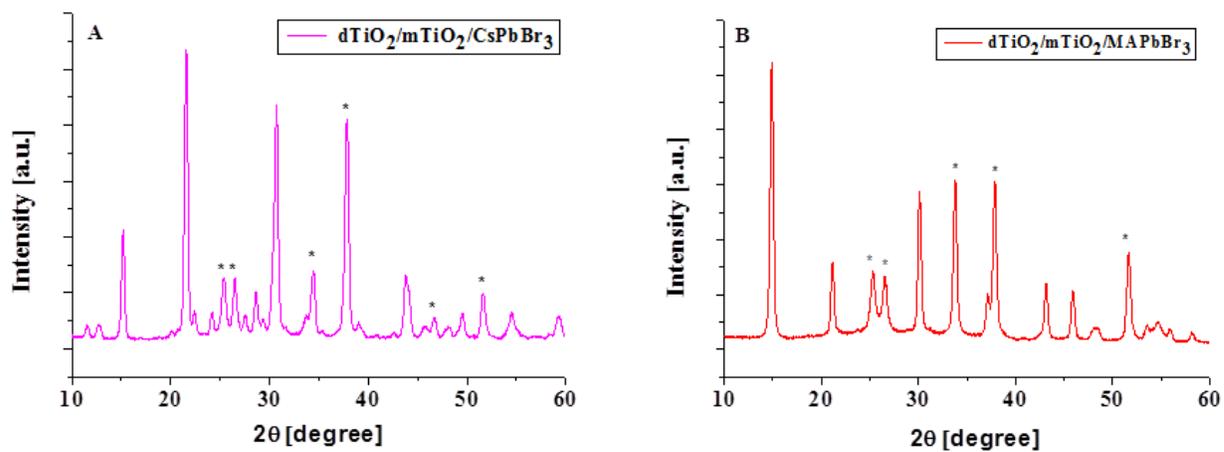

**Figure S1.** X-Ray Diffractograms of (A) $CsPbBr_3$, (B) $MAPbBr_3$ films on mp-$TiO_2$ on dense $TiO_2$ on FTO. [* indicates peaks from the d-$TiO_2$/mp-$TiO_2$ substrate].

|  | CPD (eV) | UPS (eV) |
|---|---|---|
| $MAPbBr_3$ | 4.05±0.04 | 4.15±0.1 |
| $CsPbBr_3$ | 3.85±0.02 | 3.95±0.1 |

**Table S1.** Work function values from CPD and UPS measurements.

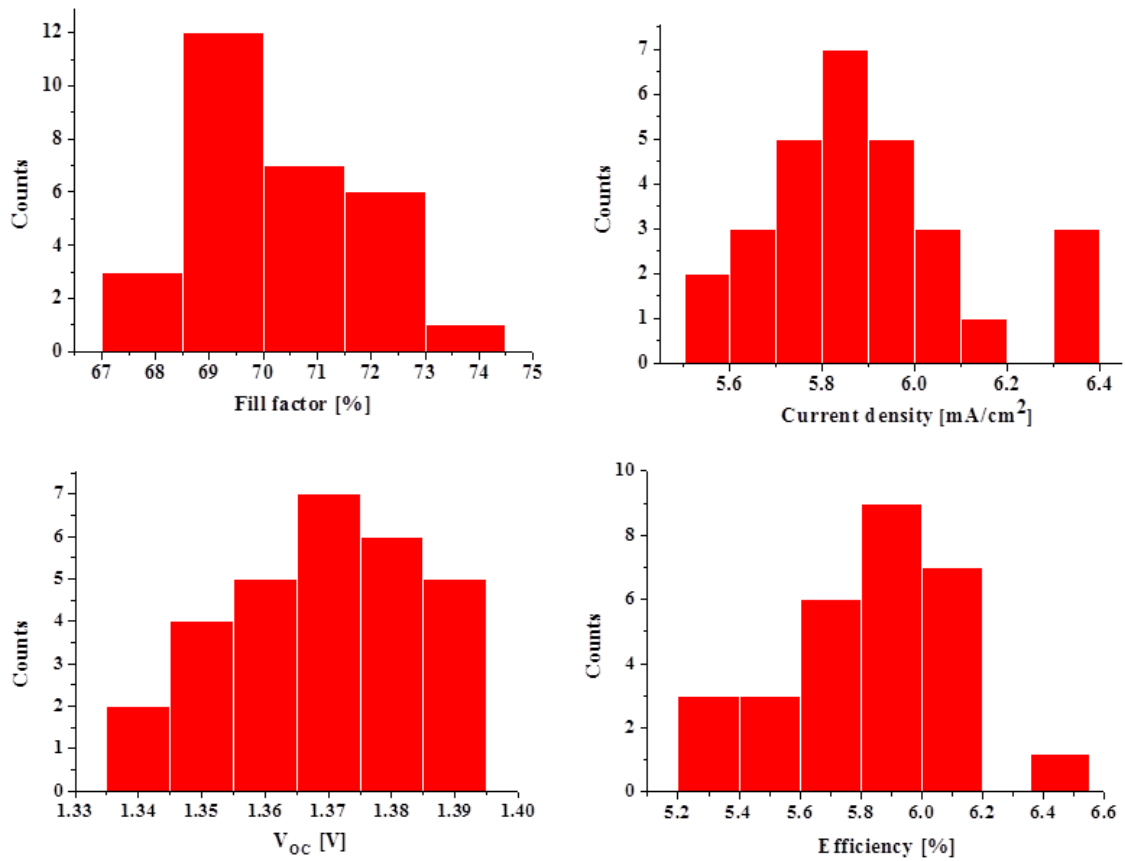

**Figure S2.** Distribution in various cell parameters for a set of 29 cells based on MAPbBr$_3$ (averaged between forward and reverse scans).

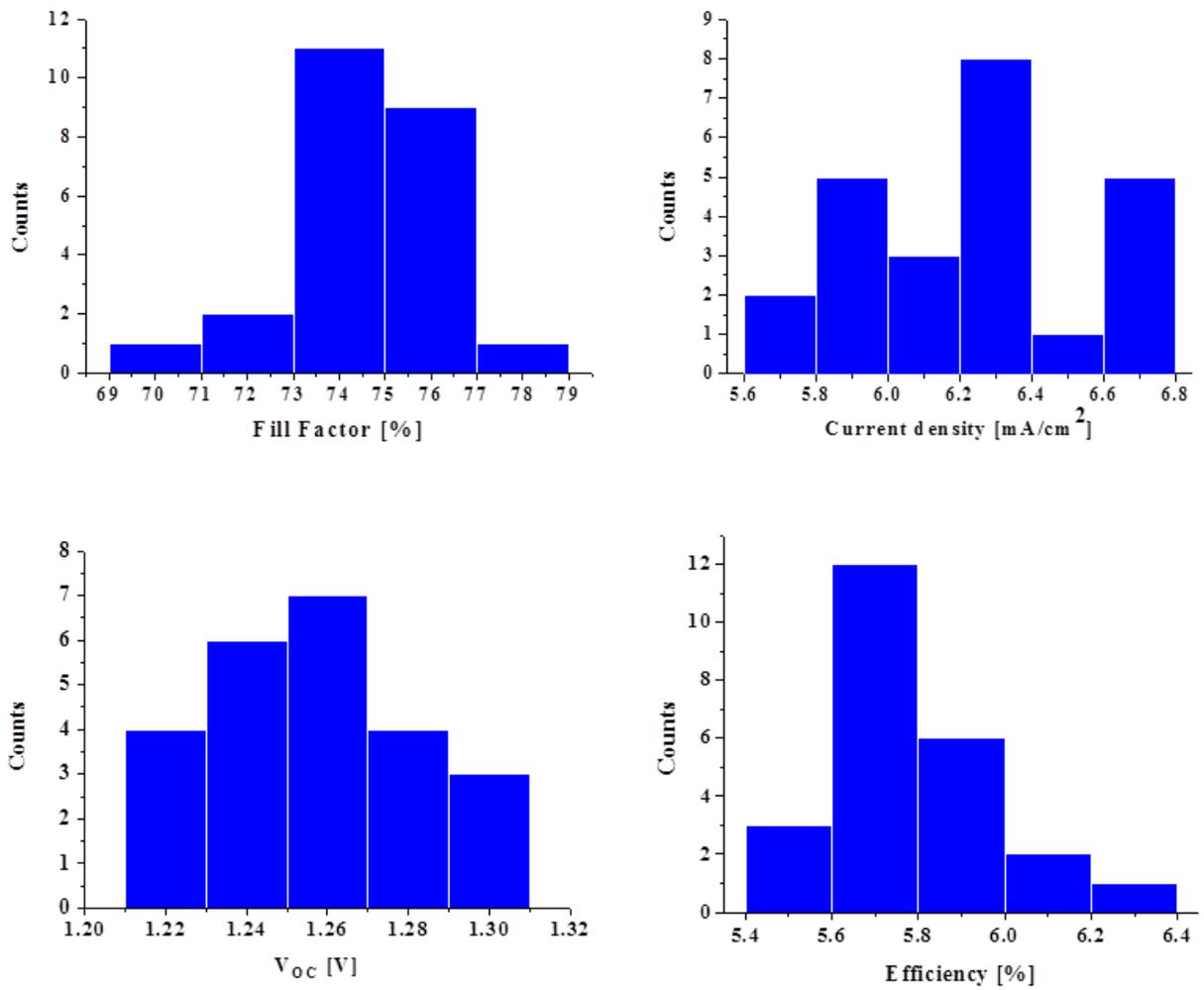

**Figure S3.** Distribution in various cell parameters for a set of 24 cells based on CsPbBr$_3$ (averaged between forward and reverse scans).

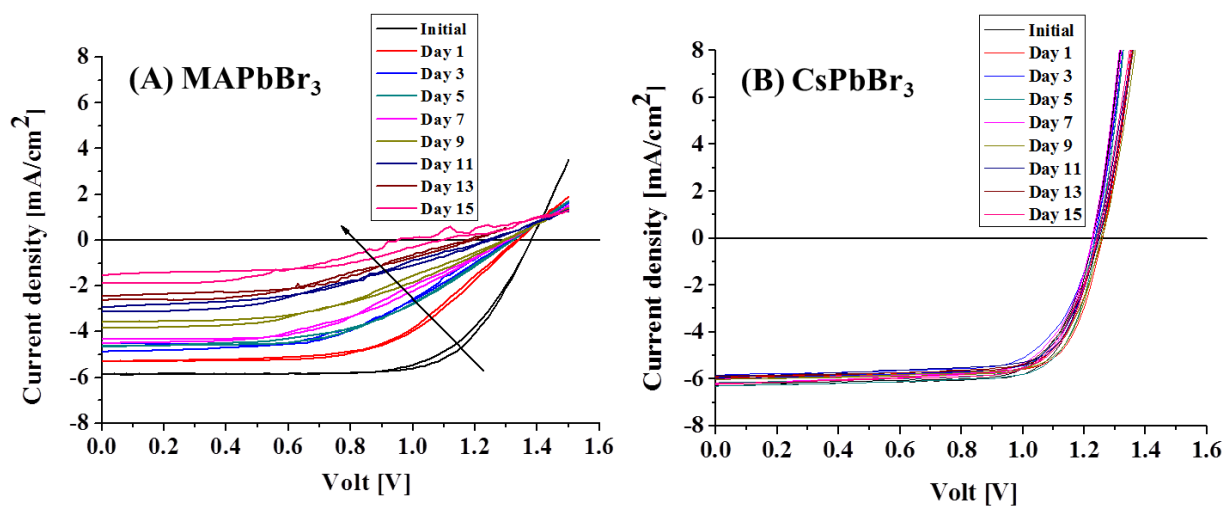

**Figure S4.** J-V curves obtained during device aging for **(A)** MAPbBr$_3$- and **(B)** CsPbBr$_3$-based devices.

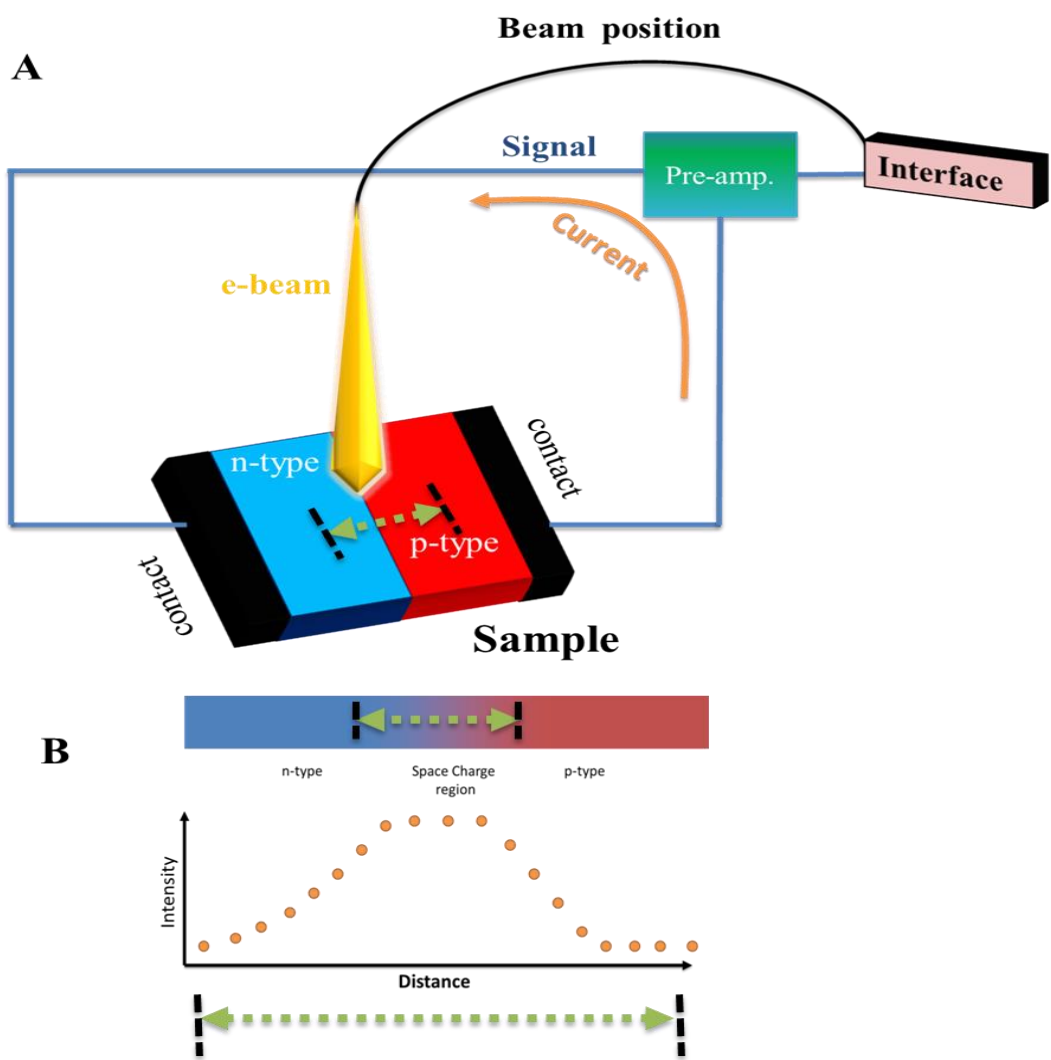

**Figure S5.** **(A)** Schematic for Electron Beam-Induced Current (EBIC) analysis, **(B)** EBIC collection profile.